\PassOptionsToPackage{unicode}{hyperref}
\PassOptionsToPackage{hyphens}{url}
\PassOptionsToPackage{dvipsnames,svgnames,x11names}{xcolor}
\documentclass[10pt,letter,onecolumn,plain]{article}
\usepackage{amsmath,amssymb}
\usepackage{lmodern}
\usepackage{iftex}
\usepackage{titlesec}
\usepackage{authblk,etoolbox}
\usepackage{tcolorbox}
\usepackage{orcidlink}
\usepackage{caption}
\usepackage[T1]{fontenc}
\usepackage[utf8]{inputenc}
\usepackage{textcomp} % provide euro and other symbols
\IfFileExists{upquote.sty}{\usepackage{upquote}}{}
\IfFileExists{microtype.sty}{% use microtype if available
  \usepackage[]{microtype}
  \UseMicrotypeSet[protrusion]{basicmath} % disable protrusion for tt fonts
}{}
\makeatletter
\@ifundefined{KOMAClassName}{% if non-KOMA class
  \IfFileExists{parskip.sty}{%
    \usepackage{parskip}
  }{% else
    \setlength{\parindent}{0pt}
    \setlength{\parskip}{6pt plus 2pt minus 1pt}}
}{% if KOMA class
  \KOMAoptions{parskip=half}}
\makeatother
\usepackage{xcolor}
\NewDocumentCommand\citeproctext{}{}
\NewDocumentCommand\citeproc{mm}{%
  \begingroup\def\citeproctext{#2}\cite{#1}\endgroup}
\makeatletter
 \let\@cite@ofmt\@firstofone
 \def\@biblabel#1{}
 \def\@cite#1#2{{#1\if@tempswa , #2\fi}}
\makeatother
\newlength{\cslhangindent}
\newlength{\csllabelwidth}
\newenvironment{CSLReferences}[2] % #1 hanging-indent, #2 entry-spacing
 {\begin{list}{}{%
  \setlength{\itemindent}{0pt}
  \setlength{\leftmargin}{0pt}
  \setlength{\parsep}{0pt}
  \ifodd #1
   \setlength{\leftmargin}{\cslhangindent}
   \setlength{\itemindent}{-1\cslhangindent}
  \fi
  \setlength{\itemsep}{#2\baselineskip}}}
 {\end{list}}
\usepackage{calc}

\usepackage[bidi=default]{babel}
\babelprovide[main,import]{american}
\def\languageshorthands#1{}
\IfFileExists{bookmark.sty}{\usepackage{bookmark}}{\usepackage{hyperref}}
\IfFileExists{xurl.sty}{\usepackage{xurl}}{} % add URL line breaks if available
\usepackage[top=2.0cm, bottom=2.0cm, right=3.5cm, left=3.5cm,
            headheight=2.2cm, reversemp, includemp, marginparwidth=0.0cm]{geometry}
\usepackage{fancyhdr}
\fancyheadoffset[L,R]{0pt} % No offset
\fancyfootoffset[L,R]{0pt} % No offset

\titleformat{\section}
  {\normalfont\sffamily\Large\bfseries}
  {}{0pt}{}
\titleformat{\subsection}
  {\normalfont\sffamily\large\bfseries}
  {}{0pt}{}
\titleformat{\subsubsection}
  {\normalfont\sffamily\bfseries}
  {}{0pt}{}
\titleformat*{\paragraph}
  {\sffamily\normalsize}
\hypersetup{
  pdftitle={GRAPE.jl: Gradient Ascent Pulse Engineering in Julia},
  pdfauthor={Michael H. Goerz, Sebastián C. Carrasco, Alastair Marshall,
Vladimir S. Malinovsky},
  pdflang={en-US},
  colorlinks=true,
  linkcolor={Maroon},
  filecolor={Maroon},
  citecolor={Blue},
  urlcolor={Blue},
  pdfcreator={LaTeX via pandoc}}

\title{GRAPE.jl: Gradient Ascent Pulse Engineering in Julia}

\author[1%
  ]{Michael~H.~Goerz%
    \,\orcidlink{0000-0003-2839-9976}\,%
    }
\author[1%
  ]{Sebastián~C.~Carrasco%
    \,\orcidlink{0000-0002-6512-9695}\,%
    }
\author[2%
  ]{Alastair~Marshall%
    \,\orcidlink{0000-0001-7772-2173}\,%
    }
\author[1%
  ]{Vladimir~S.~Malinovsky%
    \,\orcidlink{0000-0002-0243-9282}\,%
    }

\affil[1]{DEVCOM Army Research Laboratory, United States%
  }
\affil[2]{Universität Ulm, Institute for Quantum Optics, Germany%
  }
\date{}
\usepackage{xpatch}
\xapptocmd{\maketitle}{\thispagestyle{fancy}}{}{}

\begin{document}
\maketitle

\section{Summary}\label{summary}

\href{https://github.com/JuliaQuantumControl/GRAPE.jl}{The
\texttt{GRAPE.jl} package} implements Gradient Ascent Pulse Engineering
(\citeproc{ref-KhanejaJMR2005}{Khaneja et al., 2005}), a widely used
method of quantum optimal control (\citeproc{ref-BrifNJP2010}{Brif et
al., 2010}; \citeproc{ref-BrumerShapiro2003}{Brumer \& Shapiro, 2003};
\citeproc{ref-SolaAAMOP2018}{Sola et al., 2018}). Its purpose is to find
``controls'' that steer a quantum system in a particular way. This is a
prerequisite for next-generation quantum technology
(\citeproc{ref-DowlingPTRSA2003}{Dowling \& Milburn, 2003}), such as
quantum computing (\citeproc{ref-NielsenChuang2000}{Nielsen \& Chuang,
2000}) or quantum sensing (\citeproc{ref-DegenRMP2017}{Degen et al.,
2017}). For example, in quantum computing with superconducting circuits
(\citeproc{ref-KochPRA2007}{Koch et al., 2007}), the controls are
microwave pulses injected into the circuit in order to realize logical
operations on the quantum states of the system (e.g.,
\citeproc{ref-GoerzNPJQI2017}{Goerz et al., 2017}).

The quantum state of a system can be described numerically by a complex
vector \(\vert \Psi(t) \rangle\) that evolves under a differential
equation of the form

\begin{equation}\label{eq:tdse}
i \hbar \frac{\partial \vert\Psi(t)\rangle}{\partial t} = \hat{H}(\epsilon(t)) \vert\Psi(t)\rangle\,,
\end{equation}

where \(\hbar\) is the
\href{https://en.wikipedia.org/wiki/Planck_constant}{reduced Planck
constant} and \(\hat{H}\) is a matrix whose elements depend in some way
on the control function \(\epsilon(t)\). We generally know the initial
state of the system \(\vert\Psi(t=0)\rangle\) and want to find an
\(\epsilon(t)\) that minimizes some real-valued functional \(J\) that
depends on the state at some final time \(T\), as well as running costs
on \(\vert\Psi(t)\rangle\) and values of \(\epsilon(t)\) at intermediate
times. A common example is the square-modulus of the overlap with a
target state.

The defining feature of the GRAPE method is that it considers
\(\epsilon(t)\) as piecewise constant, i.e., as a vector of values
\(\epsilon_n\), for the \(n\)'th interval of the time grid. This allows
solving \autoref{eq:tdse} for each time interval, and deriving an
expression for the gradient \(\partial J / \partial \epsilon_n\) of the
optimization functional with respect to the values of the control field.
It results in an efficient numerical scheme for evaluating the full
gradient (\citeproc{ref-GoerzQ2022}{Goerz et al., 2022, fig. 1(a)}). The
scheme extends to situations where the functional is evaluated on top of
\emph{multiple} propagated states \(\{\vert \Psi_k(t) \rangle\}\) with
an index \(k\), and multiple controls \(\epsilon_l(t)\), resulting in a
vector of values \(\epsilon_{nl}\) with a double-index \(nl\). Once the
gradient has been evaluated, in the original formulation of GRAPE
(\citeproc{ref-KhanejaJMR2005}{Khaneja et al., 2005}), the values
\(\epsilon_{nl}\) would then be updated by taking a step with a fixed
step width \(\alpha\) in the direction of the negative gradient, to
iteratively minimize the value of the optimization functional \(J\). In
practice, the gradient can also be fed into an arbitrary gradient-based
optimizer, and in particular a quasi-Newton method like L-BFGS-B
(\citeproc{ref-LBFGSB.jl}{Qi \& contributors, 2022};
\citeproc{ref-ZhuATMS1997}{Zhu et al., 1997}). This results in a
dramatic improvement in stability and convergence
(\citeproc{ref-FouquieresJMR2011}{Fouquières et al., 2011}), and is
assumed as the default in \texttt{GRAPE.jl}. Gradients of the time
evolution operator can be evaluated to machine precision following
Goodwin \& Kuprov (\citeproc{ref-GoodwinJCP2015}{2015}). The GRAPE
method could also be extended to a true Hessian of the optimization
functional (\citeproc{ref-GoodwinJCP2016}{Goodwin \& Kuprov, 2016}),
which would be in scope for future versions of \texttt{GRAPE.jl}.

\section{Statement of Need}\label{statement-of-need}

There have been a number of implementations of the GRAPE method in
different contexts. GRAPE was originally developed and adopted in the
NMR community, e.g., as part of \texttt{SIMPSON}
(\citeproc{ref-TosnerJMR2009}{Tošner et al., 2009}) in C, and later as
part of \texttt{Spinach} (\citeproc{ref-HogbenJMR2011}{Hogben et al.,
2011}) and \texttt{pulse-finder} (\citeproc{ref-pulse-finder}{Ryan \&
contributors, 2013}) in Matlab. More recent implementations in Python,
geared towards more general purposes like quantum information, are found
as part of the \texttt{QuTIP} library
(\citeproc{ref-JohanssonCPC2013}{Johansson et al., 2013}), \texttt{C3}
(\citeproc{ref-WittlerPRA2021}{Wittler et al., 2021}), \texttt{QuOCS}
(\citeproc{ref-RossignoloCPC2023}{Rossignolo et al., 2023}), and
\texttt{QuanEstimation} (\citeproc{ref-ZhangPRR2022}{Zhang et al.,
2022}). The implementation of \texttt{GRAPE.jl} is also inspired by
earlier work in the \texttt{QDYN} library in Fortran
(\citeproc{ref-QDYN}{2025}). \texttt{GRAPE.jl} exploits the unique
strengths of the Julia programming language
(\citeproc{ref-BezansonSIREV2017}{Bezanson et al., 2017}) to avoid
common shortcomings in existing implementations.

As a compiled language geared towards scientific computing, Julia
delivers numerical performance similar to that of Fortran, while
providing much greater flexibility due to the expressiveness of the
language. The numerical cost of the GRAPE method is dominated by the
cost of evaluating the time evolution of the quantum system.
\texttt{GRAPE.jl} delegates this to efficient piecewise-constant
propagators in \texttt{QuantumPropagators.jl}
(\citeproc{ref-QuantumPropagators.jl}{Goerz \& contributors, 2025b}) or
the general-purpose \texttt{DifferentialEquations.jl} framework
(\citeproc{ref-RackauckasJORS2017}{Rackauckas \& Nie, 2017}).

\texttt{GRAPE.jl} builds on the concepts defined in
\texttt{QuantumControl.jl} (\citeproc{ref-QuantumControl.jl}{Goerz \&
contributors, 2025a}) to allow functionals that depend on an arbitrary
set of ``trajectories'' \(\{ \vert \Psi_k(t) \rangle\}\), each evolving
under a potentially different \(\hat{H}_k\). In contrast to the common
restriction to a single state \(\vert\Psi\rangle\) or a single unitary
\(\hat{U}\) as the dynamical state, this enables ensemble optimization
for robustness against noise (e.g., \citeproc{ref-GoerzPRA2014}{Goerz,
Halperin, et al., 2014}). The optimization over multiple trajectories is
parallelized. This makes the optimization of quantum gates more
efficient, by tracking the logical basis states instead of the gate
\(\hat{U}(t)\). Each \(\hat{H}_k\) may depend on an arbitrary number of
controls \(\{\epsilon_l(t)\}\) in an arbitrary way, going beyond the
common assumption of linear controls,
\(\hat{H} = \hat{H}_0 + \epsilon(t) \hat{H}_1\).

Julia's core feature of
\href{https://www.youtube.com/watch?v=kc9HwsxE1OY}{multiple dispatch}
allows the user to define custom, problem-specific data structures with
performance-optimized linear algebra operations. This gives
\texttt{GRAPE.jl} great flexibility to work with any custom data
structures for quantum states \(\vert \Psi_k(t) \rangle\) or the
matrices \(\hat{H}_k(\{\epsilon_l(t)\})\), and enables a wide range of
applications, from NMR spin systems to superconducting circuits or
trapped atoms in quantum computing, to systems with spatial degrees of
freedom (e.g., \citeproc{ref-DashAVSQS2024}{Dash et al., 2024}). This
also includes open quantum systems, as the structure of
\autoref{eq:tdse} holds not just for the standard Schrödinger equation,
but also for the Liouville equation, where \(\vert\Psi_k\rangle\) is
replaced by a (vectorized) density matrix and \(\hat{H}\) becomes a
Liouvillian super-operator (\citeproc{ref-GoerzNJP2014}{Goerz, Reich, et
al., 2014}).

The rise of machine learning generated considerable interest in using
the capabilities of frameworks like Tensorflow
(\citeproc{ref-Tensorflow}{Abadi et al., 2016}), PyTorch
(\citeproc{ref-PaszkeNIPS2019}{Paszke et al., 2019}), or JAX
(\citeproc{ref-JAX}{Bradbury et al., 2018}) for automatic
differentiation (AD) (\citeproc{ref-Griewank2008}{Griewank \& Walther,
2008}) to evaluate the gradient of the optimization functional. This has
the benefit that it allows for arbitrary functionals
(\citeproc{ref-AbdelhafezPRA2019}{Abdelhafez et al., 2019},
\citeproc{ref-AbdelhafezPRA2020}{2020};
\citeproc{ref-LeungPRA2017}{Leung et al., 2017},
\citeproc{ref-quantum-optimal-control}{2021}). In contrast, the GRAPE
method and all of its existing implementations are formulated only for a
``standard'' set of functionals that essentially measure the overlap of
a propagated state with a target state. Unfortunately, AD comes with a
large numerical overhead that makes the method impractical. Goerz et al.
(\citeproc{ref-GoerzQ2022}{2022}) introduced the use of ``semi-automatic
differentiation'' that limits the numerical cost to exactly that of the
traditional GRAPE scheme. It does this by employing AD only for the
evaluation of the derivative
\(\partial J/\partial \langle \Psi_k(T) \vert\), and only if that
derivative cannot be evaluated analytically. \texttt{GRAPE.jl} is built
on the resulting generalized GRAPE scheme. As necessary, it can use any
available AD framework in the Julia ecosystem to enable the minimization
of non-analytical functionals, such as entanglement measures
(\citeproc{ref-GoerzPRA2015}{Goerz et al., 2015};
\citeproc{ref-WattsPRA2015}{Watts et al., 2015}).

\section{Acknowledgements}\label{acknowledgements}

Research was sponsored by the Army Research Laboratory and was
accomplished under Cooperative Agreement Numbers W911NF-23-2-0128 (MG)
and W911NF-24-2-0044 (SC). The views and conclusions contained in this
document are those of the authors and should not be interpreted as
representing the official policies, either expressed or implied, of the
Army Research Laboratory or the U.S. Government. The U.S. Government is
authorized to reproduce and distribute reprints for Government purposes
notwithstanding any copyright notation herein.

\section*{References}\label{references}
\addcontentsline{toc}{section}{References}

\phantomsection\label{refs}

\end{document}